\documentclass[preprint2]{aastex631}
\let\bfseries\mdseries 
\shorttitle{Combining SOSS and G395H}
\shortauthors{Liu, Wang, Rustamkulov, \& Sing}
\usepackage{amsmath}
\usepackage{multirow}
\let\tablenum\relax
\usepackage{siunitx}
\graphicspath{{./}{mainFigures/}}

\begin{document}

\title{Rereduction and Calibration of JWST NIRSpec and NIRISS Commissioning Data on HAT-P-14 b with the Latest Methods}

\title{Unveiling the atmosphere of the super-Jupiter HAT-P-14 b with JWST NIRISS and NIRSpec}

\author[0000-0003-0685-3525]{Rongrong Liu}
\email{rongrong.liu@cfa.harvard.edu
}
\altaffiliation{These authors contributed equally to this work.}

\affiliation{Center for Astrophysics $\vert$ Harvard \& Smithsonian, 60 Garden Street, Cambridge, MA 02138, USA}

\affiliation{William H. Miller III Department of Physics \& Astronomy, Johns Hopkins University, 3400 N Charles St, Baltimore, MD 21218, USA}

\author[0000-0002-6379-3816]{Le-Chris Wang}
\email{lwang178@jhu.edu}
\altaffiliation{These authors contributed equally to this work.}
\affiliation{William H. Miller III Department of Physics \& Astronomy, Johns Hopkins University, 3400 N Charles St, Baltimore, MD 21218, USA}

\author[0000-0003-4408-0463]{Zafar Rustamkulov}
\affiliation{Department of Earth and Planetary Science, Johns Hopkins University, 3400 N. Charles Street, Baltimore, MD 21218, USA}

\author[0000-0001-6050-7645]{David K. Sing}
\affiliation{William H. Miller III Department of Physics \& Astronomy, Johns Hopkins University, 3400 N Charles St, Baltimore, MD 21218, USA}
\affiliation{Department of Earth and Planetary Science, Johns Hopkins University, 3400 N. Charles Street, Baltimore, MD 21218, USA}



\begin{abstract}
\noindent We report the combined JWST NIRSpec/G395H and NIRISS/SOSS transmission spectrum of the transiting super-Jupiter HAT-P-14 b, from 0.60 $\mu m$ to 5.14 $\mu m$. Initial analysis of these data reported a near-featureless spectrum at NIRSpec wavelengths range (2.87 $\mu m$ to 5.14 $\mu m$) consistent with the small atmospheric scale height of the planet and unexplained bumps and wiggles at NIRISS wavelengths range (0.6 $\mu m$ to 2.8 $\mu m$). Here, we produce a self-consistent  spectrum of HAT-P-14 b's atmosphere with an up-to-date reduction. We detect H$_2$O (3.09 $\sigma$) both across NIRISS/SOSS wavelengths range and at the bluest end of NIRSpc/G395H as well as a gray cloud deck (1.90 $\sigma$). We constrain the atmospheric metallicity of HAT-P-14 b to be roughly Solar, with [Fe/H] $= -0.08^{+0.89}_{-0.98}$, consistent with the planet mass-metallicity relationship. The differences compared to previous works are likely due to the improved STScI \verb|jwst| pipeline, which highlights the need to reanalyze the early NIRISS/SOSS transiting exoplanet targets with the latest methods. As HAT-P-14 b is placed as the 805th best target for transmission spectroscopy according to Transmission Spectroscopy Metrics (TSM), our results showcase JWST's unparalleled photometric precision which can easily characterize a thousand exoplanets' atmospheres through transmission spectroscopy.
\end{abstract}



\section{Introduction} \label{sec:intro}

The unparalleled photometric precision and infrared capability of the James Webb Space Telescope (JWST) have enabled us to study  exoplanet atmospheres at wavelengths never probed before with unprecedented precision. Through transmission spectroscopy,  \citep{2000ApJ...537..916S, 2001ApJ...553.1006B}, JWST has discovered a rich repository of chemical species in exoplanet atmospheres previously inaccessible \citep{2017PASP..129f4501B} and provided insights into planets' internal structure, formation, and evolution (e.g., \citealt{2023Natur.614..649J, 2023Natur.614..653A, 2023Natur.614..659R,2023Natur.614..664A,2023Natur.614..670F, 2024Natur.630..831S, 2024arXiv240303325B}). In particular, the Near Infrared Spectrograph (NIRSpec) G395H mode \citep{2022A&A...661A..83B} and the Near Infrared Imager and Slitless Spectrograph (NIRISS) Single Object Slitless Spectroscopy (SOSS) mode \citep{2023PASP..135g5001A} are two workhorse modes for exoplanet atmosphere observations. NIRSpec/G395H provides high-resolution (R $\sim$ 2700) spectroscopy from 2.87 $\mu m$ to 5.14 $\mu m$, which enables coverage of multiple carbon species such as CH$_4$, CO$_2$, and CO as well as photochemical byproducts such as SO$_2$ for the first time. NIRISS/SOSS covers 0.6 $\mu m$ to 2.85 $\mu m$ at medium resolution (R $\sim$ 700), enabling us to measure multiple near-infrared H$_2$O bands, alkali metals such as Na and K, the presence of hazes, and potentially the carbon-bearing species at the reddest wavelengths. The synergy of NIRISS/SOSS and NIRSpec/G395H  provides information-rich insights into exoplanet atmospheres. 


As part of the JWST commissioning program, the transiting exoplanet HAT-P-14 b \citep{2010ApJ...715..458T} was observed by both NIRSpec/G395H (PID 1118; PI: Proffitt; \citealt{2023PASP..135a8002E}, E23 hereafter) and NIRISS/SOSS (PID 1541; PI: Espinoza; \citealt{2023PASP..135g5001A}, A23 hereafter). It is a dense (3.44 $M_J$; 1.42 $R_J$), short-period (4.6 days) planet around a bright ($J$ = 9.09), low-activity F-star \citep{2017AJ....153..136S}. Because it is massive, its scale height is relatively small ($H\sim$ 150 km), which results in only small features (a few tens of ppm) in its transmission spectrum. Therefore, this planet is an excellent target for commissioning purposes, as a relatively featureless spectrum would be expected, and any wavelength-dependent variations are most likely because of instrumental effects. Despite this, water features are expected to be present in the transmission spectrum of HAT-P-14 b based on simulations (A23), with amplitudes up to 50 ppm. Initial analysis of both datasets showed some discrepancies. An analysis of the NIRSpec/G395H dataset for HAT-P-14 b was presented by E23, who concluded that the transmission spectrum is featureless. On the other hand, an analysis of the NIRISS/SOSS counterpart was presented by A23, who reported a transmission spectrum that was inconsistent with being flat. In particular, a rising slope towards the bluest wavelengths that deviates from the flat line with amplitudes of at least 100 ppm was present, and a "break" at $1.4~\mu m$ could not be explained.

In this work, we reanalyze both NIRSpec/G395H and NIRISS/SOSS datasets for HAT-P-14 b with the latest methods to thoroughly investigate whether atmospheric features are present. We combine both datasets and present a panchromatic spectrum, which we fit with atmospheric models. 
We describe the observations and our data reduction steps in Section \ref{sect:obs} and detail the light curve analysis in Section \ref{sect:lightcurve}. We present the atmospheric retrieval setup and results in Section \ref{sect:results}. We summarize and discuss our findings in Section \ref{sect:discussion} and conclude in Section \ref{sec:conclusion}.

\section{Observations and Data Reductions}\label{sect:obs}

\subsection{Observation Outline}
\subsubsection{NIRSpec G395H}
The first HAT-P-14 b transit event was observed during the JWST commissioning campaign with NIRSpec/BOTS mode and G395H grating on UTC May 30, 2022, from 00:30 to 07:02 (PID 1118; PI: Proffitt). The 6-hour-long observation includes 1139 integrations with 20 NRSRAPID groups per integration, covering the wavelength range from 2.87 $\mu$m to 5.14 $\mu$m. The S1600A1 aperture and the SUB2048 subarray were used, and data was collected by both NRS1 and NRS2 detectors (E23). We downloaded the uncalibrated data of this observation from the MAST archive. The dataset DOI is \dataset[10.17909/am3x-rn73]{http://dx.doi.org/10.17909/am3x-rn73}.

\subsubsection{NIRISS SOSS}
Observation of HAT-P-14 b with NIRISS SOSS was made during commissioning (program ID: 1541, PI. Espinoza). The TSO started on UTC June 08, 2022, and observed one transit for 6.1 hours or 572 NISRAPID integrations with 6 groups per integration. A standard GR700XD/CLEAR combination was used, together with SUBSTRIP256 to capture the diffraction of all three spectral orders. For our analysis, we downloaded the publicly available uncalibrated data from the MAST archive. The dataset DOI is \dataset[10.17909/f37r-sp75]{http://dx.doi.org/10.17909/f37r-sp75}.

\subsection{Data Reduction}
\subsubsection{NIRSpec G395H}
We reduce the NIRSpec G395H dataset using the Fast InfraRed Exoplanet Fitting for Lightcurves (FIREFLy) pipeline \citep{2022ApJ...928L...7R, 2023Natur.614..659R}. The data reduction steps of Firefly can be divided into two stages. 

Stage 1 performs corrections at the detector level. It is based on Stage 1 of the \texttt{jwst} Python calibration pipeline developed by STScI \citep{2023zndo...6984365B} with the TSO configuration.  We include a custom $1/f $ subtraction step after the superbias step in the \texttt{jwst} pipeline to perform de-striping at the group level. We set the rows of top 7 and bottom 6 pixels to be the background reference pixels used in the pipeline. The dark-current step and the jump step in the \texttt{jwst} pipeline are both skipped. For the custom group-level 1/f subtraction step, we create a mask for the point spread function (PSF) in the image with a threshold flux $ = 3.6DN/s$. We then subtract the column median value of pixels outside the PSF mask from the image and repeat this process for all groups in the data. 

In Stage 2 of Firefly, we perform spatial and temporal cleaning by replacing bad pixels and cosmic rays. The $1/f$ subtraction step described above is applied again at the integration level to ensure all remaining 2D background noise is cleaned. We then calculate a 1D wavelength array by first interpolating the 2D calibration wavelength map to obtain a wavelength value for every pixel and then getting the value at the trace center for each column. We measure the shift of the spectral trace for both detectors with cross-correlation and shift-stabilize the spectrum with flux-conserving cubic spline interpolation. \textbf{The measured positional shifts are small (on the order of $10^{-3}$ pixels, or 0.1 mas, for both detectors) and random, showing no systematic drifts. Applying this correction reduces position-dependent systematics, consistent with previous JWST analyses \citep{2024Natur.630..831S, 2023Natur.614..659R}. }
This shift-stabilization process is repeated twice to ensure that the spectrum's position-dependent trends are fully removed. For both detectors, \textbf{we extract the 1D time-series data for light curve fitting using an aperture designed to get the best spectrophotometric data. We use a constant aperture width across all wavelengths with a modified box extraction that sums pixel values within $\pm n\sigma$ from the trace center and including fractional pixels at the aperture edges.  Here, $\sigma$ is the median PSF core width determined by fitting a Gaussian profile to the trace at each wavelength, and the aperture width factor $n$ is set to 3.8 for G395H.}

\subsubsection{NIRISS SOSS}
We use Fast InfraRed Exoplanet Fitting for Lightcurves (FIREFLy) \citep{2022ApJ...928L...7R, 2023Natur.614..659R} for our reduction. While FIREFLy was originally written for reducing NIRSpec-PRISM and G395H time series observations, we have updated and optimized it to work on NIRISS-SOSS dataset with almost the same workflow as that for NIRSpec. 

Like our reduction for NIRSpec, the reduction for SOSS is composed of three stages. Stage 1 of FIREFLy mostly follows the STScI pipeline \citep{2023zndo...6984365B} \footnote{\href{https://jwst-pipeline.readthedocs.io/en/latest/index.html}{https://jwst-pipeline.readthedocs.io/en/latest/index.html}}, with the exception of 1/f subtraction and saturation correction. 



For saturation correction, we find that the default STScI step is insufficient to solve the problem in the HAT-P-14 b observation. The resulting transmission spectra show an excessive and unexpected point-to-point scatter in the saturated regions at a magnitude of thousands of ppm, an issue that was also observed in \cite{2023Natur.614..659R}. Indeed, the HAT-P-14 b observation produced saturation beginning at NGROUP = 3 and affect all the pixels between 0.9 and 1.5 $\mu m$ of order 1 by NGROUP = 6 (A23). To correct the saturation problem, we manually flag the whole columns corresponding to saturated pixels starting at NGROUP = 4 and progressively flag more columns as saturated ``up-the-ramp"; only the groups before the saturation flags are used for slope fitting. We avoid flagging saturated pixels at NGROUP = 3 to alleviate biases that could be introduced in the ramp-fit step due to the first group effect \citep{2024NatAs.tmp..128C}.

For $1/f$ correction, FIREFLy treats $1/f$ noise at the group level rather than at the integration level as $1/f$ noise is time-varying and could be different from group to group. To correct for $1/f$ noise, we take the temporal median through a given number of exposures for each group in each integration. We then take the column-wise median of the difference between each group frame and its temporally local median image and subtract it from that frame. 

\begin{figure*}[t]
    \plotone{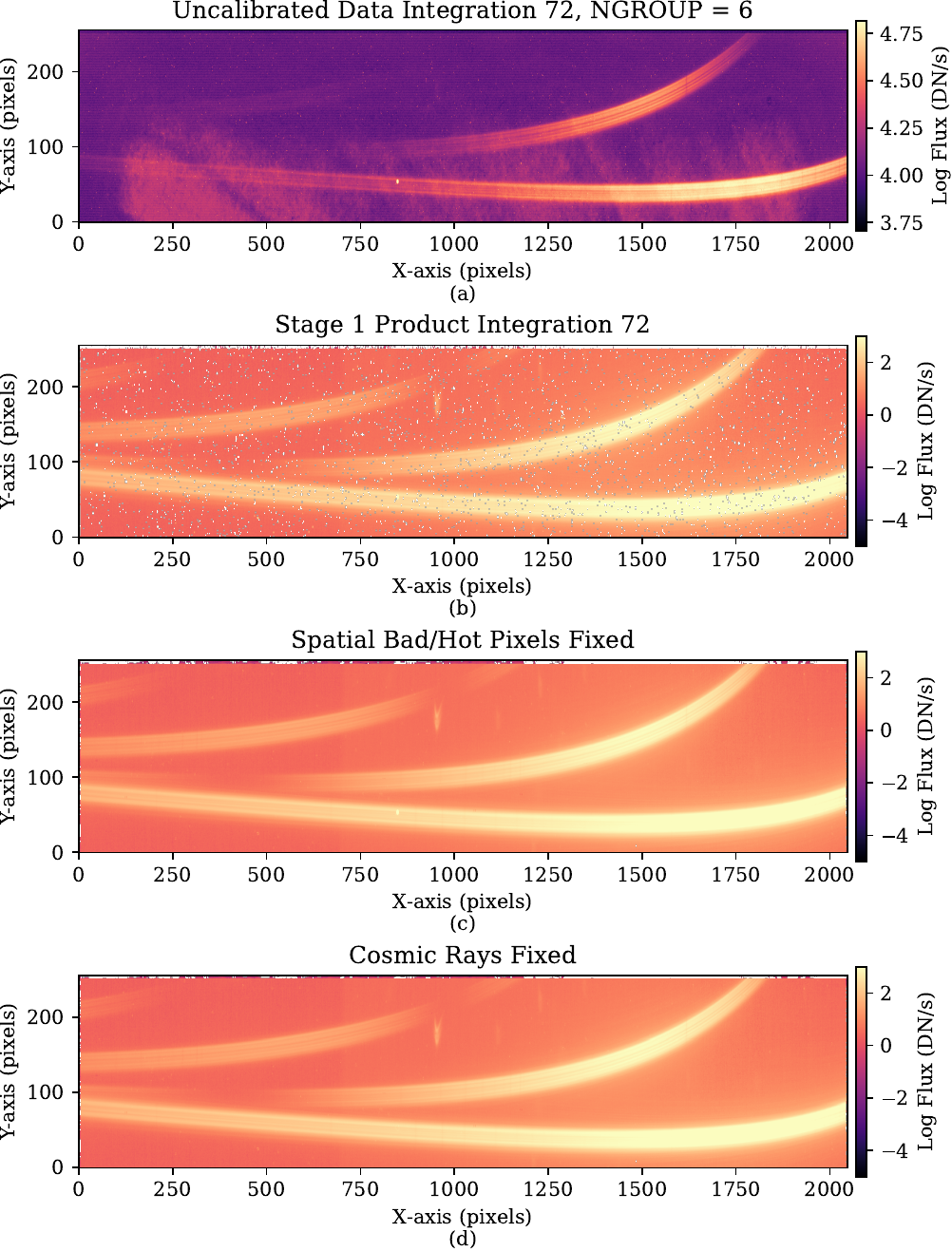}
    \caption{Products of FIREFLy reduction for the NIRISS-SOSS HAT-P-14 b dataset at different steps up till cosmic rays removal. \textbf{(a)}: Raw uncalibrated frame in Data Numbers (DN). \textbf{(b)}: After stage 1 calibration. \textbf{(c)}: After bad and hot pixels correction. \textbf{(d)}: After cosmic rays removal.}
    \label{fig:cosmic}
\end{figure*}

Stage 2 of FIREFLy is composed of bad pixel cleaning, background removal, and 1D extraction. We clean the pixels with bad-quality flags and the remaining hot pixels by median-filtering each frame and fixing those bad pixels to the median of the neighboring pixels. We then remove cosmic rays by both using the L.A.Cosmic algorithm \textbf{\citep{van_Dokkum_2001}} and rejecting the temporal outliers. We show images of the 72nd integration to illustrate the products of each step we describe in Figure \ref{fig:cosmic}.

\begin{figure*}[t]
    \plotone{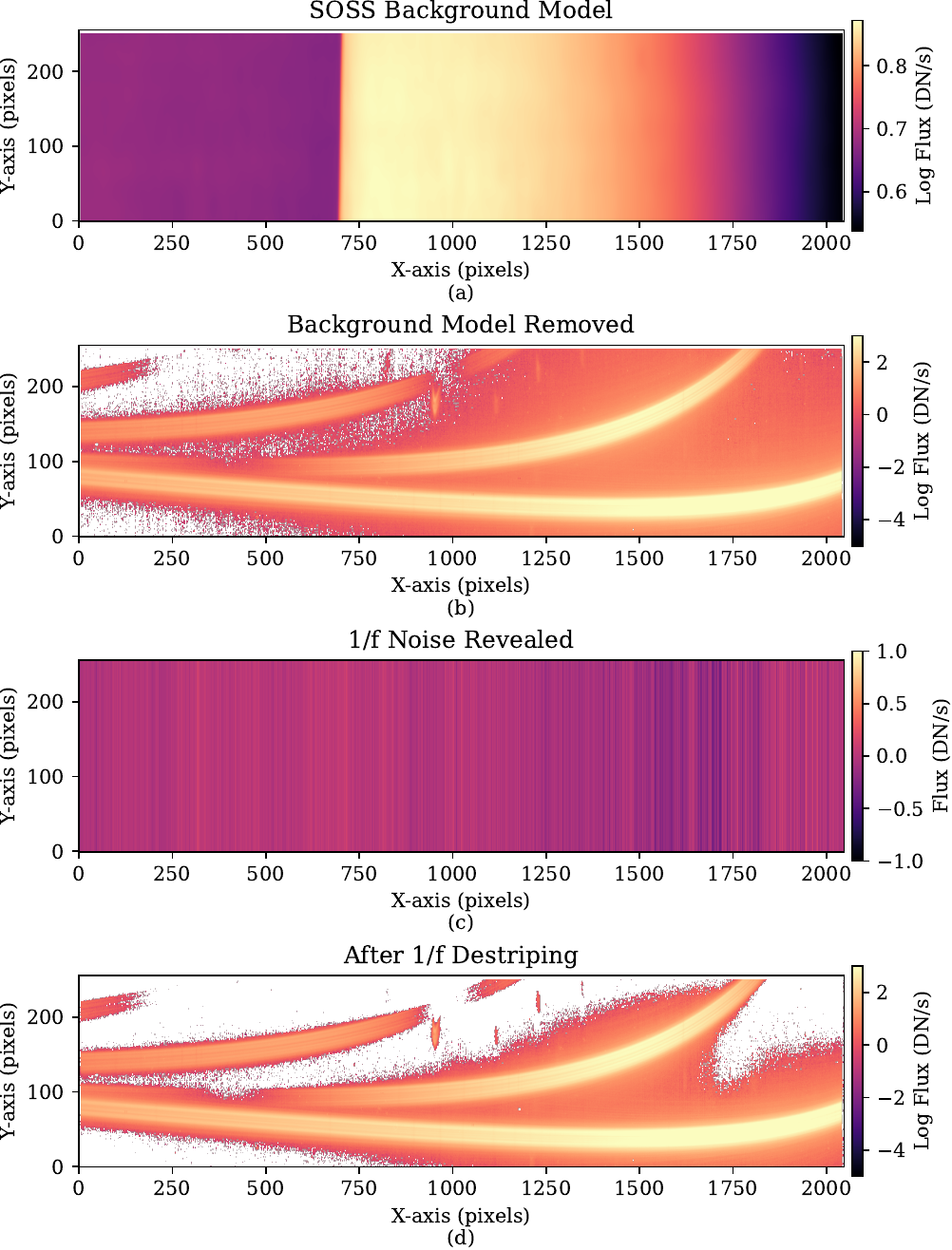}
    \caption{1/f noise and background removal steps of FIREFLy. \textbf{(a)}: The background model for SOSS measured during commissioning; a jump of flux levels due to the zodiacal light can be seen at x $\sim$ 700. \textbf{(b)}: The same frame as Figure \ref{fig:cosmic} after background subtraction. \textbf{(c)}: The 1/f noise shown in the frame. \textbf{(d)}: The same frame as (b) after 1/f destripping.}
    \label{fig:background}
\end{figure*}

The SOSS background has a unique ``jump" in flux level at $\sim 2.1 ~ \mu m$ (panel (a) of Figure \ref{fig:background}) due to zodiacal light falling off the JWST ``pick-off" mirror; this is apparent in Figure \ref{fig:cosmic}. The zodiacal light also causes a wavelength-dependent variation towards the bluer wavelengths. If not handled correctly, the resulting transmission spectra could be heavily diluted \citep{2023MNRAS.524..835R}. To remove the background, FIREFLy subtracts a model SOSS background created during commissioning which we download from \verb|jwst-docs|\footnote{\href{https://jwst-docs.stsci.edu/}{https://jwst-docs.stsci.edu/}}. We mask out the PSF and scale the background to the flux level of the remaining region that is almost unaffected by the spectral trace and the zeroth order contamination. We assume the zodiacal background to be time-invariant in the 6.1 hours TSO, so we subtract the same scaled background model from all the frames. We show a background-removed frame in the panel (b) of Figure \ref{fig:background}.

To fully remove $1/f$ noise, FIREFLy again applies the same temporal median method as that in stage 1 at the integration level. We then mask the stellar point spread function and subtract the median unmasked pixel values of each column. This helps remove both the $1/f$ noise and any remaining background. We created the point spread function mask carefully so that the spectral traces (including most of their wings) and prominent zeroth order contaminants are included. The product from this step is shown in panel (d) of Figure \ref{fig:background}.

\begin{figure*}[t]
    \plotone{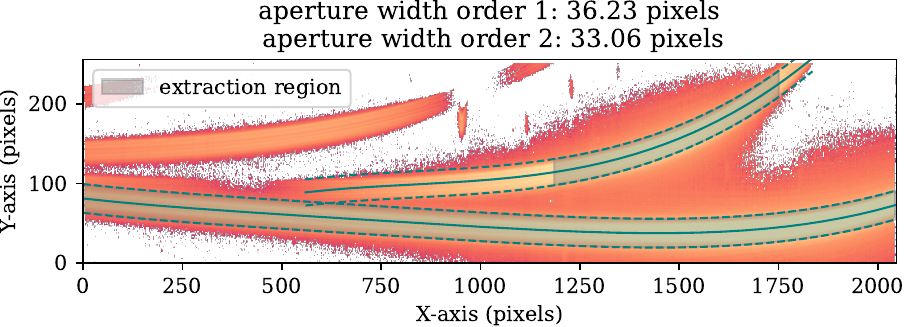}
    \caption{Spectrophotometry extraction step of FIREFLy. We choose aperture widths for each spectral order so that the out-of-transit scatter for the transit light curve is minimized. We extract the spectrophotometry bounded by the boxed regions.}
    \label{fig:aperture}
\end{figure*}

FIREFLy uses cross-correlation to measure the positional shift of the spectral traces and align all the frames. We obtain the wavelength map \verb|jwst-niriss-wavemap-0021.fits| available from the JWST Calibration Reference Data System (CRDS) as our wavelength solution. We extract the spectrophotometry for order 1 and order 2 separately with an optimized constant aperture width that minimizes the scatter of the off-transit data, \textbf{similar to how we process the NIRSpec G395H data. The aperature width factor $n$ is set to 15.17 for SOSS order 1, and 11.65 for SOSS order 2.} The spectrophotometry extraction for both orders is shown in Figure \ref{fig:aperture}.

\section{Light Curve Analysis}\label{sect:lightcurve}

\subsection{NIRSpec G395H}
We integrate the observed stellar light flux across the entire wavelength range to obtain the white light curve for the two NIRSpec detectors NRS1 and NRS2. The data is trimmed in both time and wavelength dimensions to exclude the effects of known bad images. The first 20 integrations are excluded in the time trimming. The wavelength trimming is different for the two detectors: for NRS1, the first 575 and last 8 indices in the wavelength array are trimmed; for NRS2, the first 5 and last 18 indices in the wavelength array are trimmed. 


After obtaining the white light curve, we use its out-of-transit part to determine what systematics to include in the model by running linear regression with different systematics combinations. The best model is chosen based on the Bayesian Information Criterion (BIC), which balances the model complexity with the goodness of fit. For both detectors, the best systematic model includes only a first-order linear term, $lin1$. 
We then fit the white light curve using transit model generated by Python \verb|batman| package \citep{2015PASP..127.1161K} to obtain the orbital parameters. The orbital period $p$, argument of pariastron $\omega$, and the eccentricity $e$ are fixed to the literature values from \cite{2017A&A...602A.107B} and \cite{2017AJ....153..136S} in the white light curve fit. Their values are shown in Table \ref{table_params}. In the initial fit, we set the scaled semi-major axis $a/R_*$, the impact parameter $b$, and the variation in mid-transit time, $t_0$ (the difference between the actual mid-transit time and the time corresponding to the minimum flux level during transit), to free parameters. Then we jointly fit the other 5 parameters: mean flux $f_0$, scaled planetary radius $R_p/R_*$, the linear term $lin1$, and the two quadratic limb darkening coefficients $q_1$ and $q_2$ as defined by Equation \ref{ldc} in \cite{2013MNRAS.435.2152K}.
\begin{equation}
    q_1 = (u_1+u_2)^2 \quad q_2 = \frac{u_1}{2(u_1+u_2)}
    \label{ldc}
\end{equation}

\begin{table*}
\centering
\caption{Best-fitting white light curve parameters}
\label{table_params}
    \footnotesize
    \begin{tabular}{cccc}
        \hline
        \hline
        Parameter & Value & Description & Reference \\
        \hline
        $P$ & 4.62767 (fixed) & Orbital period (days) & \cite{2017AJ....153..136S} \\
        \hline
        $\omega$ & 106.1 (fixed) & Argument of periastron ($^\circ$) & \cite{Bonomo_2017}  \\
        \hline
        e & 0.11 (fixed) & Eccentricity & \cite{Bonomo_2017}  \\
        \hline
        \textbf{[M/H]} & \textbf{0.110 (fixed)} & \textbf{Stellar metallicity (dex)} & \textbf{\cite{Bonomo_2017} } \\
        \hline
        \textbf{$\log g$} & \textbf{4.25 (fixed) } & \textbf{Stellar surface gravity ($\log_{10}$(\si{cm/s^{2}}))} & \textbf{\cite{Bonomo_2017} }\\
        \hline
        \textbf{$T_\text{eff}$} & \textbf{6600} & \textbf{Stellar effective temperature (K)} & \textbf{\cite{Bonomo_2017} } \\
        \hline
        $a/R_{\ast,\text{NIRSpec}}$ & $8.111 \pm 0.031$ & NIRSpec weighted average scaled semi-major axis  & This work\\
        \hline
        $b_{\text{NIRSpec}}$ & $1.014 \pm 0.001$ & NIRSpec weighted average impact parameter & This work\\
        \hline
        $T_{0, \text{NIRSpec/NRS1}}$ & $59729.706739\pm0.000051$& NIRSpec NRS1 mid-transit time (days, MJD) & This work \\
        \hline
        $T_{0, \text{NIRSpec/NRS2}}$ &  $59729.706665\pm 0.000066$ & NIRSpec NRS2 mid-transit time (days, MJD) & This work \\
        \hline
        $T_{0, \text{NIRISS/Order1}}$ & $59738.462267\pm 0.000045$ & SOSS order 1 mid-transit time (days, MJD) & This work \\
        \hline
        $T_{0, \text{NIRISS/Order2}}$ & $59738.462252 \pm 0.000085$  & SOSS order 2 mid-transit time (days, MJD) & This work \\
        \hline
        $R_p^2/R^2_{\ast, \text{NIRSpec/NRS1}}$ & $6639\pm 56$ & NIRSpec NRS1 scaled planetary radius (ppm) & This work\\
        \hline
        $R_p^2/R^2_{\ast, \text{NIRSpec/NRS2}}$ & $6675\pm 101$ & NIRSpec NRS2 scaled planetary radius (ppm) & This work\\
        \hline
        $R_p^2/R^2_{\ast, \text{NIRISS/Order1}}$ & $6750\pm 40$ & SOSS Order 1 scaled planetary radius (ppm)& This work\\
        \hline
        $R_p^2/R^2_{\ast, \text{NIRISS/Order2}}$ & $6714\pm 68$ & SOSS Order 2 scaled planetary radius (ppm)& This work\\
        \hline
        $q_{1,\text{NIRSpec/NRS1}}$ & $0.0199\pm 0.0062$ & NIRSpec NRS1 first LDC & This work\\
        \hline
        $q_{2,\text{NIRSpec/NRS1}}$ & $0.0044\pm 0.1059$ & NIRSpec NRS1 second LDC & This work\\
        \hline
        $q_{1,\text{NIRSpec/NRS2}}$ & $0.0226\pm 0.0087$ & NIRSpec NRS2 first LDC & This work\\
        \hline
        $q_{2,\text{NIRSpec/NRS2}}$ & $0.6055\pm 0.2931$ & NIRSpec NRS2 second LDC & This work\\
        \hline
        $q_{1,\text{NIRISS/Order1}}$ & $0.2780\pm 0.0157$ & SOSS order 1 first LDC & This work\\
        \hline
        $q_{2,\text{NIRISS/Order1}}$ & $0.000\pm 0.000$ & SOSS order 1 second LDC & This work\\
        \hline
        $q_{1,\text{NIRISS/Order2}}$ & $0.4319\pm 0.0317$ & SOSS order 2 first LDC & This work\\
        \hline
        $q_{2,\text{NIRISS/Order2}}$ & $0.000\pm 0.000$ & SOSS order 2 second LDC & This work\\
        \hline
    \end{tabular}
\end{table*}

\begin{figure*}
     \plottwo{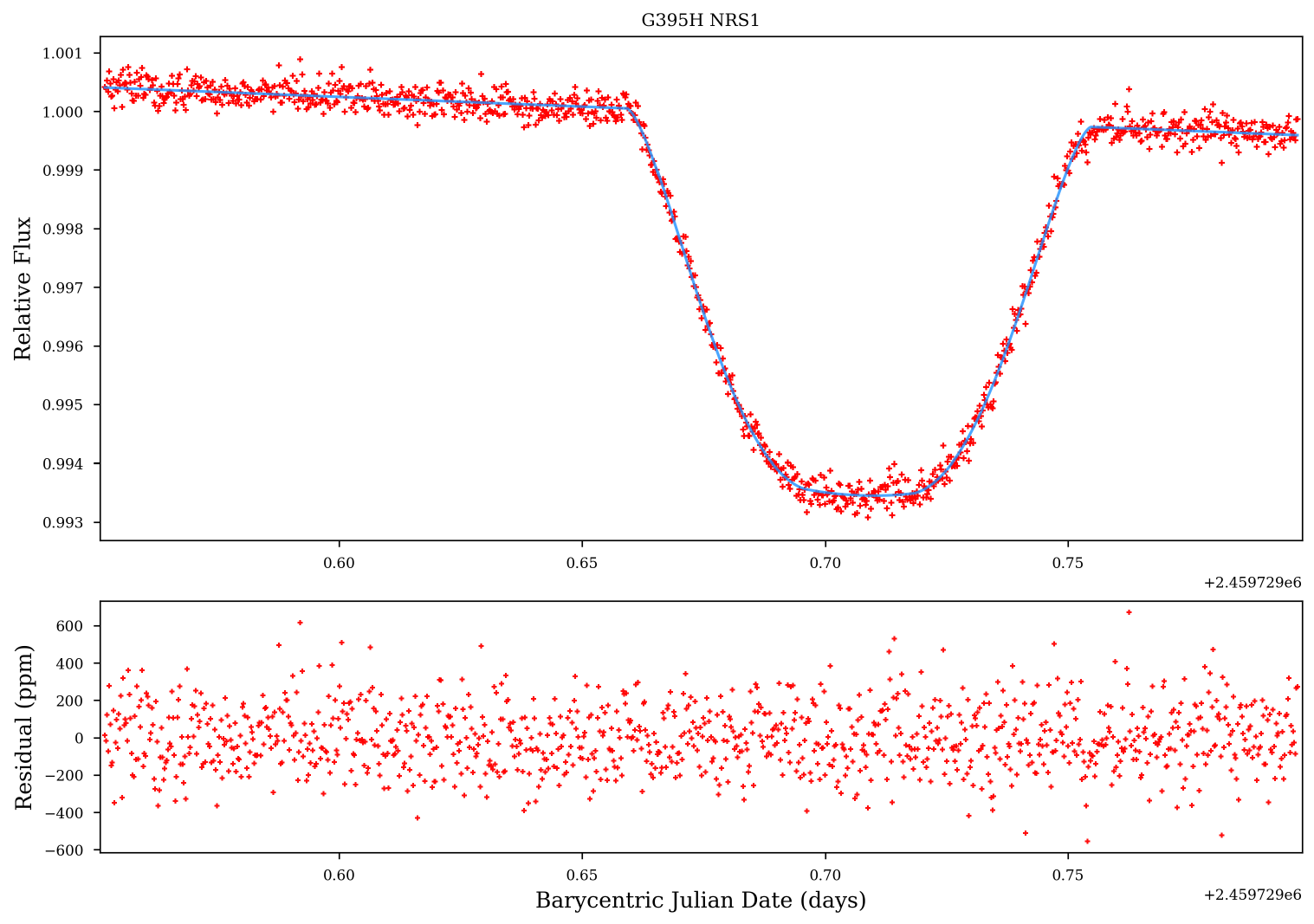}{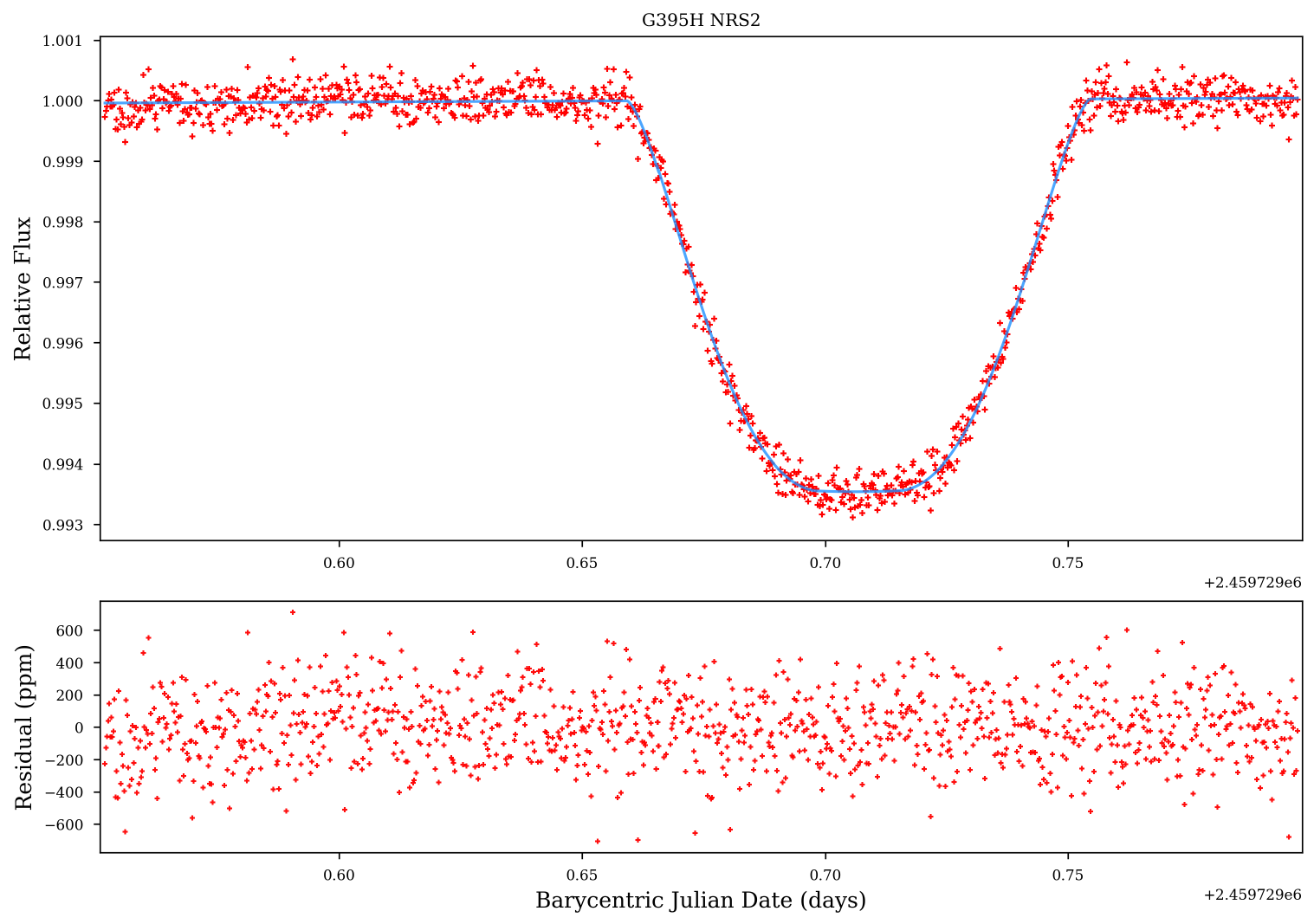}
     \caption{White light curve fits for detectors NRS1 and NRS2 of NIRSpec, along with the residuals between model and observed light curves. }
     \label{fig:wlc_fit_nirspec}
 \end{figure*}

\begin{figure*}
     \plottwo{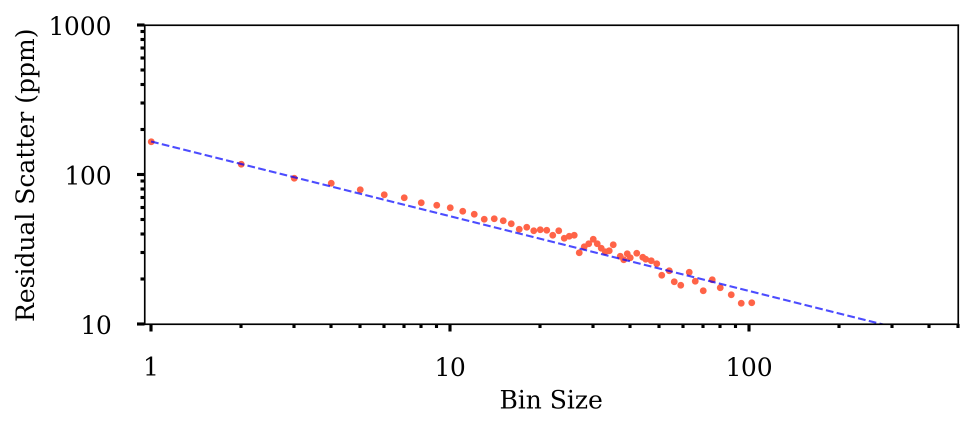}{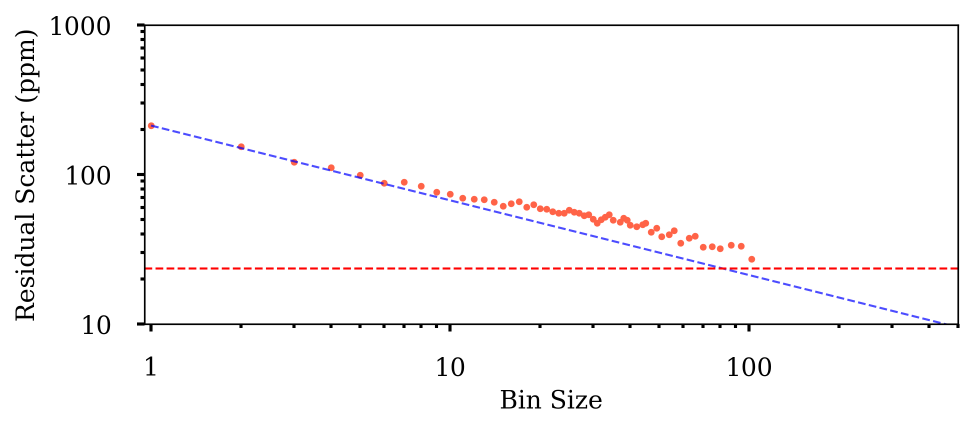}
     \label{noise_nirspec}
     \caption{Plots of the relationship between bin size and residual scatter for detectors NRS1 and NRS2 of NIRSpec. The dashed blue line is the theoretical decrease for white noise (proportional to $1/\sqrt{N}$) and the horizontal red dashed line is an estimate of the rednoise} 
 \end{figure*}

 From the results of the initial white light curve fit, the value of $a/R_*$ is $8.293\pm 0.070$ for NRS1 and $8.065\pm 0.035$ for NRS2, and the value of $b$ is $1.014\pm 0.004$ for NRS1 and $1.022\pm 0.001$ for NRS2; the best-fit values of $a/R_*$ and $b$ are calculated by taking the weighted average of the results from both detectors: $a/R_* = 8.111\pm 0.031$ and $b = 1.022 \pm 0.001$. To ensure the consistency of best-fit parameters, we then fix the values of $a/R_*$ and $b$ for both detectors and run a second fit to determine the best-fit values of the other parameters. The best-fit parameters we obtained are shown in Table \ref{table_params}. Those values are used as priors in the transmission spectrum fitting. We use the Markov Chain Monte Carlo sampler (\verb|emcee|) in the \verb|lmfit| package to perform the fit, with 1000 total steps, and get the posterior probability distribution of the parameters. The fitted white light curves and residuals for both detectors are shown in Figure \ref{fig:wlc_fit_nirspec}. We then binned the residuals in time to show the decrease of noise against binning size and to investigate the rednoise calculated from the difference between the binned standard deviation and the overall standard deviation. The noise plot for NRS1 and NRS2 is shown in Figure \ref{noise_nirspec}.

 
 To extract the transmission spectrum, we first divide the spectrum into wavelength bins adaptively so that the flux level in each bin is roughly the same. Then we use the same strategy to fit the free parameters $R_p/R_*$, $f_0$, $lin1$, $q_1$, and $q_2$, while fixing the values of $a/R_*$, $b$, and $t_0$ to the values from white light curve fit. By plotting the fitted value of $R_p/ R_*$ in wavelength bins, we obtain the planet's transmission spectrum in the combined wavelength range of both detectors. 
 From the fit results, we noticed that the values of limb-darkening coefficients, $q_1$ and $q_2$, in the NIRSpec wavelength range is close to the stellar model value, so it is reasonable to fix their values to the model.
 
 
\subsection{NIRISS SOSS}

\begin{figure*}[t]
    \plottwo{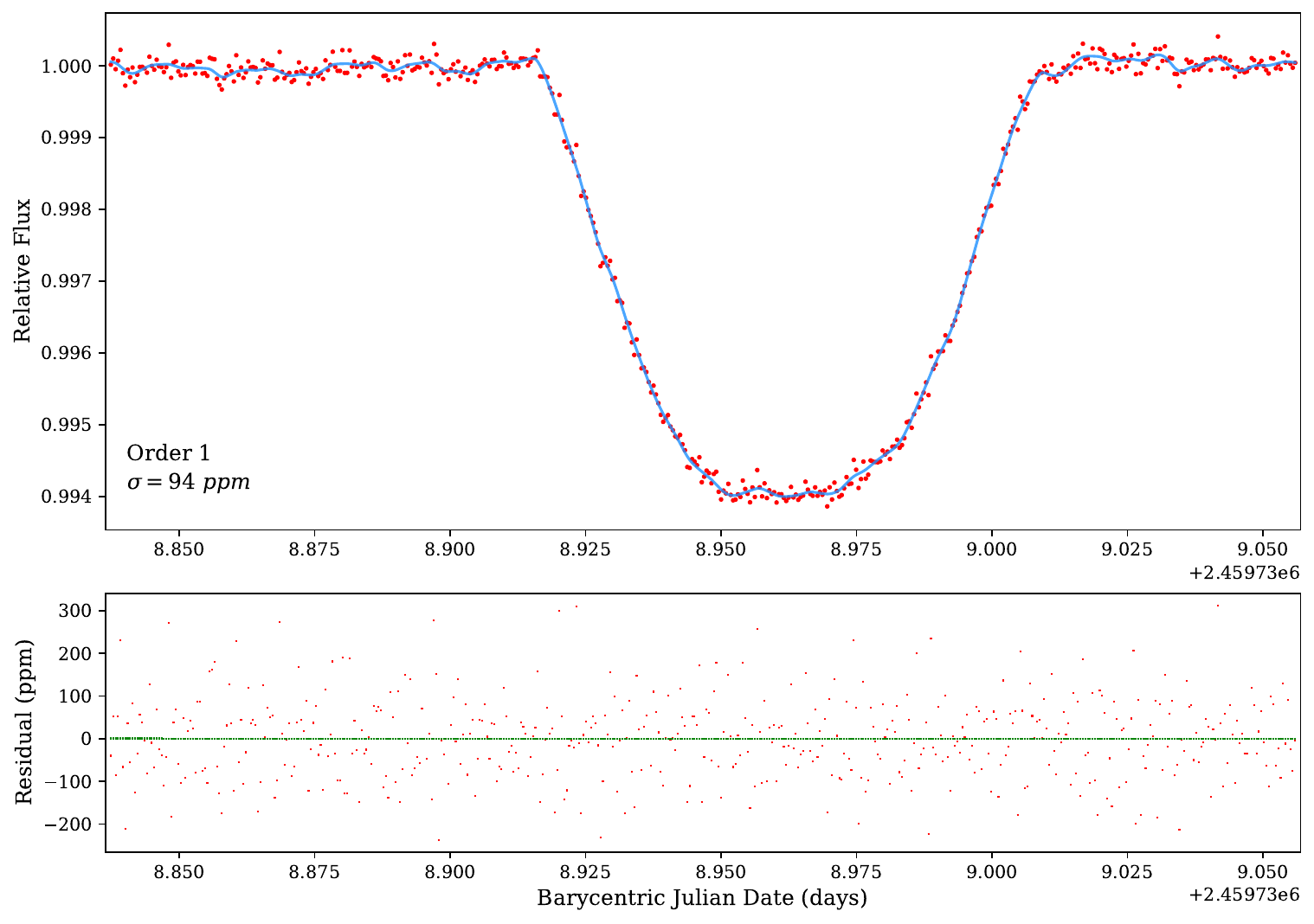}{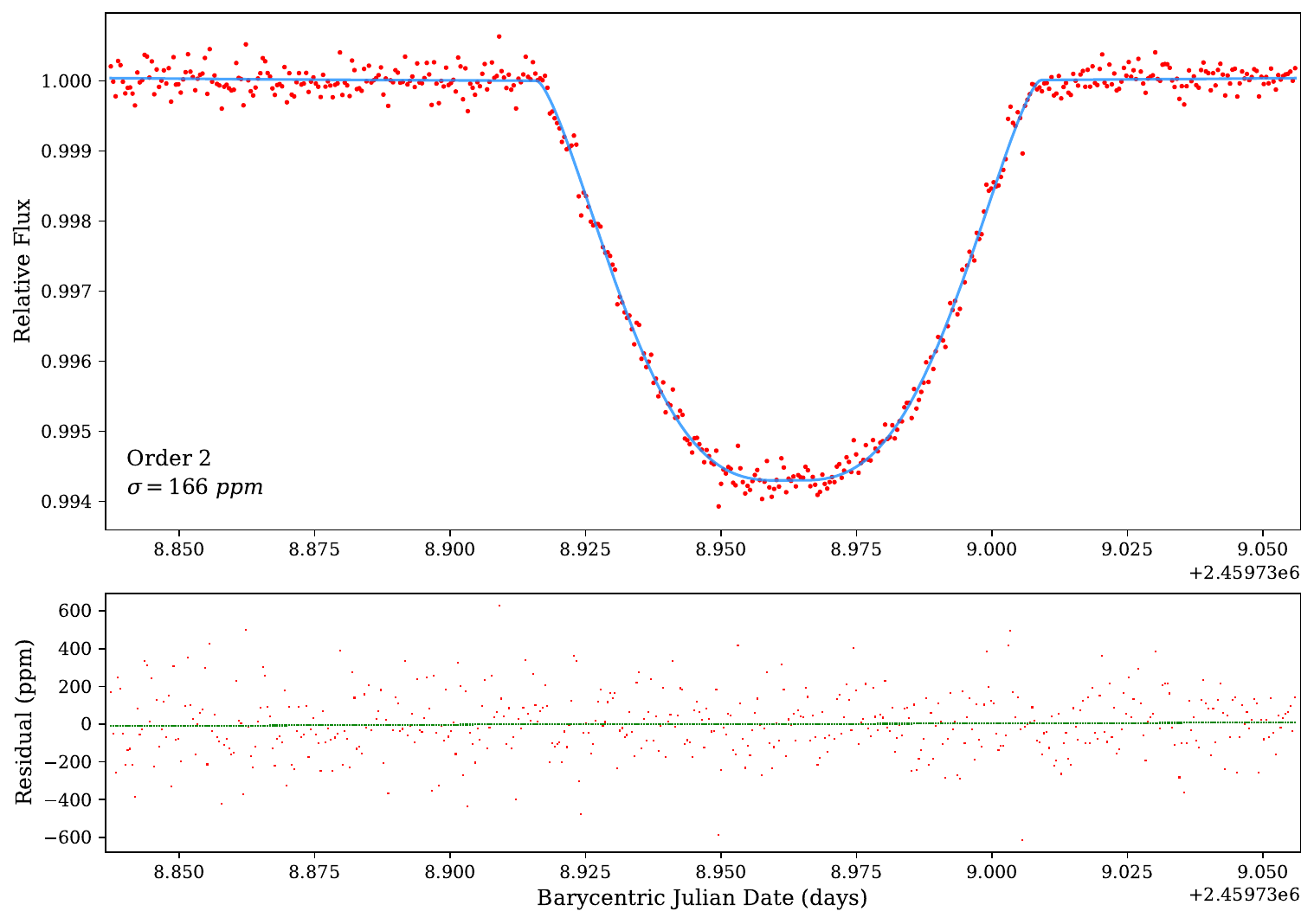}
    \caption{White light curve fits for both order 1 and order 2. \textit{Top}: the white light curves for SOSS order 1 and order 2, respectively shown on the left and right. The reduced $\chi^2$ for SOSS order 1 is 1.01 and that for SOSS order 2 is 1.02. \textit{Bottom}: The residuals for the white light curve fits. }
    \label{fig:soss_wlc}
\end{figure*}

We sum the flux in the extraction region in Figure \ref{fig:aperture} for to obtain the white light curve for each order. We exclude the first 80 integrations because they exhibit a non-linear baseline flux trend. We fit the white light curve with \verb|batman| \citep{2015PASP..127.1161K}. To ensure consistency, we fix the impact parameter, $b$, and scaled orbital semi-major axis, $a/R_{\ast}$, to the weighted average of the fitted results from NRS1 and NRS2 of the NIRSpec dataset; we also fix the orbital period, longitude of periastron, and eccentricity to the same values used in white light curve fitting in NIRSpec. We allow the following parameters to vary: the scaled planet radius, $R_{p}/R_{\ast}$, mean flux, $f_0$, two parameters of the quadratic limb darkening law, $q_1$ and $q_2$, with parametrization following \cite{2013MNRAS.435.2152K}, and the best systematics vector suggested by the bayesian information criterion (the first and second order linear terms, $lin1$ and $lin2$, in the case of order 1 and the second order linear term, $lin2$, in the case of order 2). We iteratively use the Levenberg-Marquart least-squares minimization algorithm to perform initial light curve fits and then use Markov Chain Monte Carlo (MCMC) with wide uninformative priors to explore the parameter space. Additionally, we noticed a correlated noise with unknown origin in the residual of our initial SOSS order 1 white light curve fit with a peak at about 30 min, which is also present in the reduction from the commissioning team (A23). To remove these systematics in the final transmission spectrum, we use the Gaussian-smoothed residual from the white light curve fit as a common mode detrending vector. The self-correction from the common mode would cause the residual for the white light curve fit to be underestimated, but it would allow us to detrend the spectroscopic light curves better if such noise is common to all the wavelength bins. The fitted white light curves for both orders are shown in Figure \ref{fig:soss_wlc}.


To obtain the transmission spectrum, we perform a joint fit of the light curve and the systematics to each wavelength channel. The systematics vector for order 1 is composed of $lin1$, $lin2$, and the common mode, $comm$. It is composed of $lin2$ for order 2. Because the transit is extremely grazing ($b = 1.014$), it is difficult to constrain the limb-darkening coefficients. Therefore, consistent with our treatment with the NIRSpec data, we fix both $q_1$ and $q_2$ for the quadratic limb darkening law to the theoretically derived values from the 3D stellar model \citep{2015A&A...573A..90M}. Since HAT-P-14 is an \textbf{F-type} star, we expect the actual limb-darkening coefficients to be consistent with the stellar model's predictions. We also fix $T_0$ to the fitted value from the white light curve fit and $a/R_{\ast}$ and $b$ to the fitted values from NIRSpec.

\begin{figure*}[t]
    \plottwo{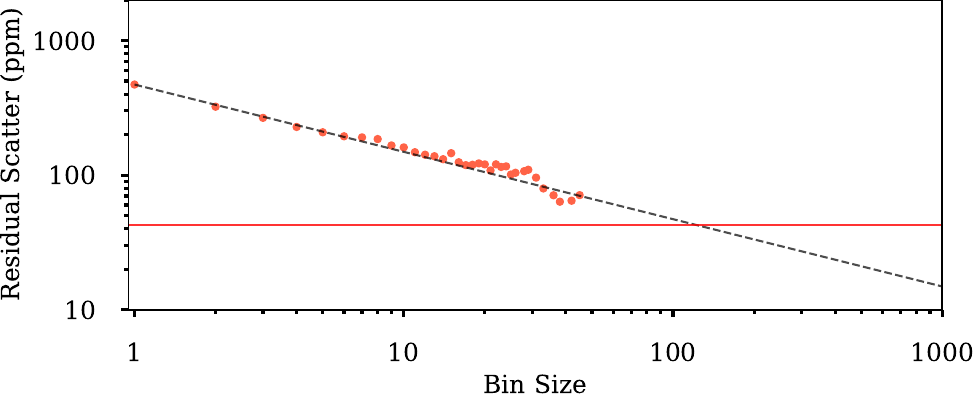}{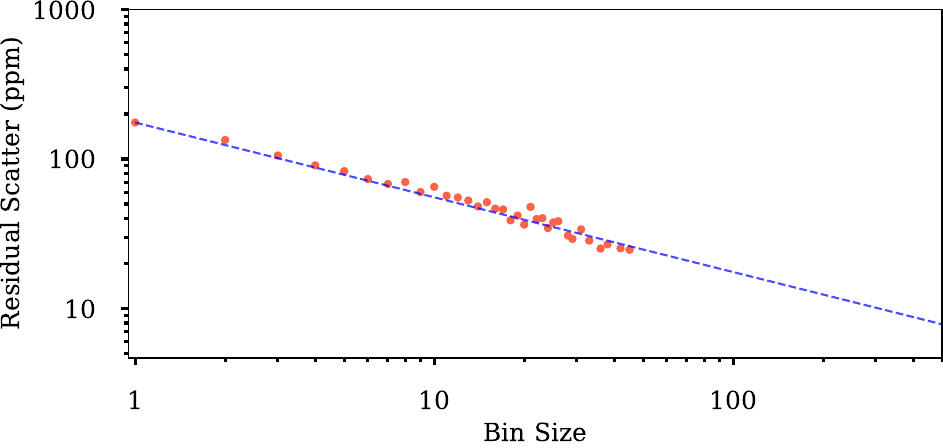}
    \caption{The relationship between bin size and residual scatter for a particular spectroscopic bin of order 1 (left) and the white light curve of order 2 (right). We have chosen not to display the same plot for the SOSS order 1 white light curve, as the use of common-mode correction results in an underestimation of the residual scatters. The dashed blue line is the theoretical decrease for white noise (proportional to $1/\sqrt{N}$) and the horizontal red dashed line is an estimate of the red noise}
    \label{fig:soss_rednoise}
\end{figure*}


\section{Results}\label{sect:results}
\subsection{Observed Transmission Spectrum}
\begin{figure*}
    \plotone{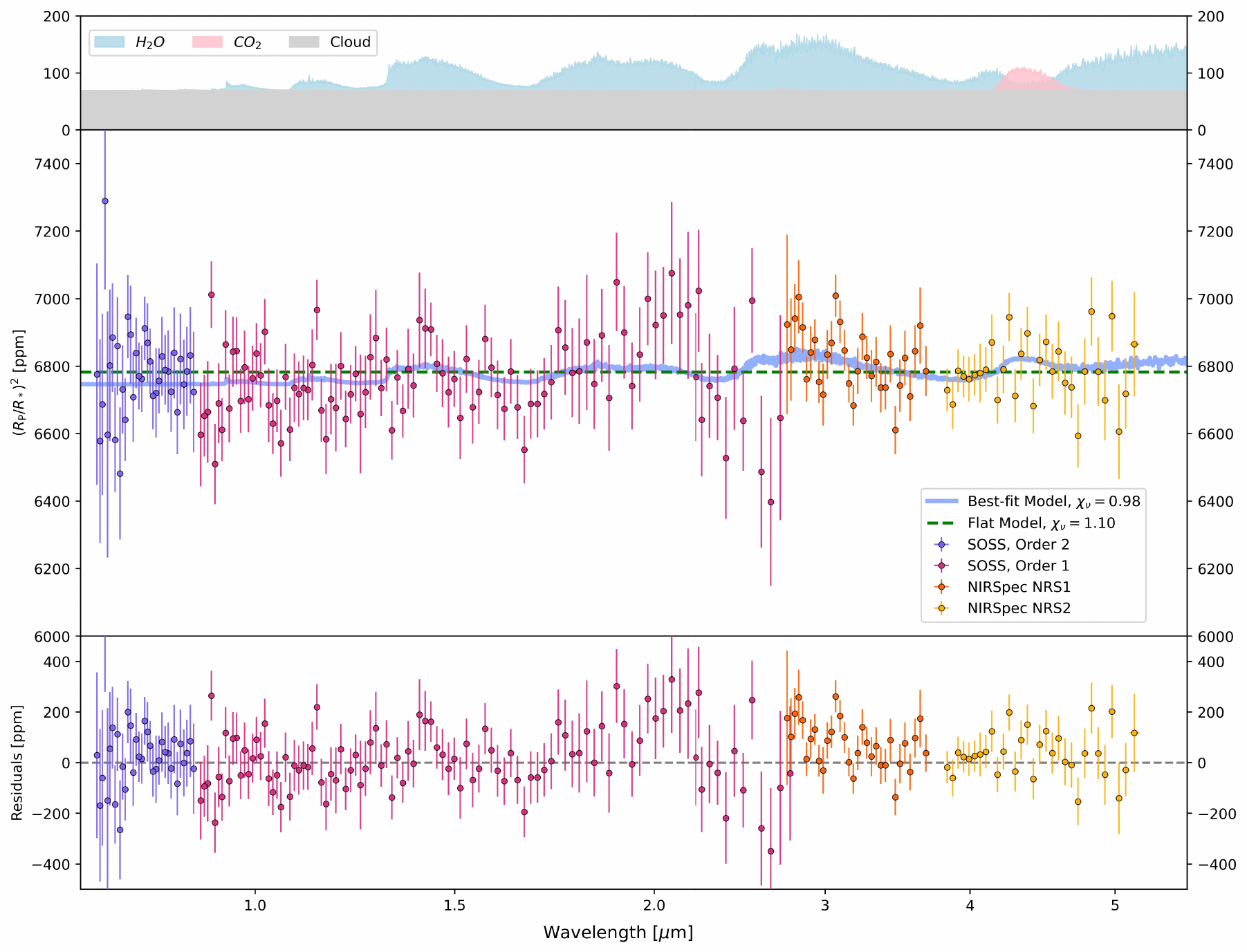}
    \caption{The observed transmission spectrum and retrieval analysis results. (a) Retrieval results showing the contributions of major atmosphere components. We excluded the contributions from $\mathrm{CO}$ and $\mathrm{CH_4}$ since they won't be visible in the plot. (b) The transmission spectrum of planet HAT-P-14 b, extracted separately from fitting the lightcurves observed by NIRISS SOSS, NIRSpec NRS1 G395H, and NIRSpec NRS2 G395H. The fitted offset of 237ppm between SOSS and NIRSpec is added to the SOSS part of the spectrum. The full model spectrum retrieved by PetitRADTRANS and the flat atmosphere model is also shown. (c) The residual spectrum obtained by subtracting the best-fit model from the observed transmission spectrum.}
    \label{fig: comb_spec}
\end{figure*}

We use the transmission spectra from fitting the light curves in the wavelength range covered by NIRSpec NRS1 and NRS2 and NIRISS SOSS to generate a combined transmission spectrum.  The combined spectrum is shown in Figure \ref{fig: comb_spec}. We calculated the weighted mean $(R_p/R_*)^2$ and the uncertainty of the three components and show them in Table \ref{tab:spec_stats}. The weighted mean of the transit depths from  NRS1 and NRS2 are consistent, while the same for SOSS is about 250 ppm lower. This significant offset exists despite using the same reduction pipeline, suggesting a systematic difference between the instruments, possibly due to the variance in their detector sensitivity. Therefore, we include an offset term for SOSS to account for this difference in the retrieval step (\ref{sec:setup}).


\begin{table}[]
\begin{centering}
\begin{tabular}{ccc}
\hline
Detector and Mode  & weighted $\sigma(R_p/R_*)^2$ [ppm]\\ \hline
NIRISS SOSS        & $6522.4\pm 9.8 $                 \\
NIRSpec NRS1 G395H & $6815.7\pm 12.3 $                    \\
NIRSpec NRS2 G395H &  $6780.8\pm 13.8$               \\ \hline
\end{tabular}

\caption{The statistics of transmission spectra as obtained from NIRISS SOSS, NIRSpec NRS1 G395H, and NIRSpec NRS2}
\label{tab:spec_stats}
\end{centering}

\end{table}

\subsection{Binning Test}

\begin{figure}[t]
    \centering
    \includegraphics[width=1\columnwidth]{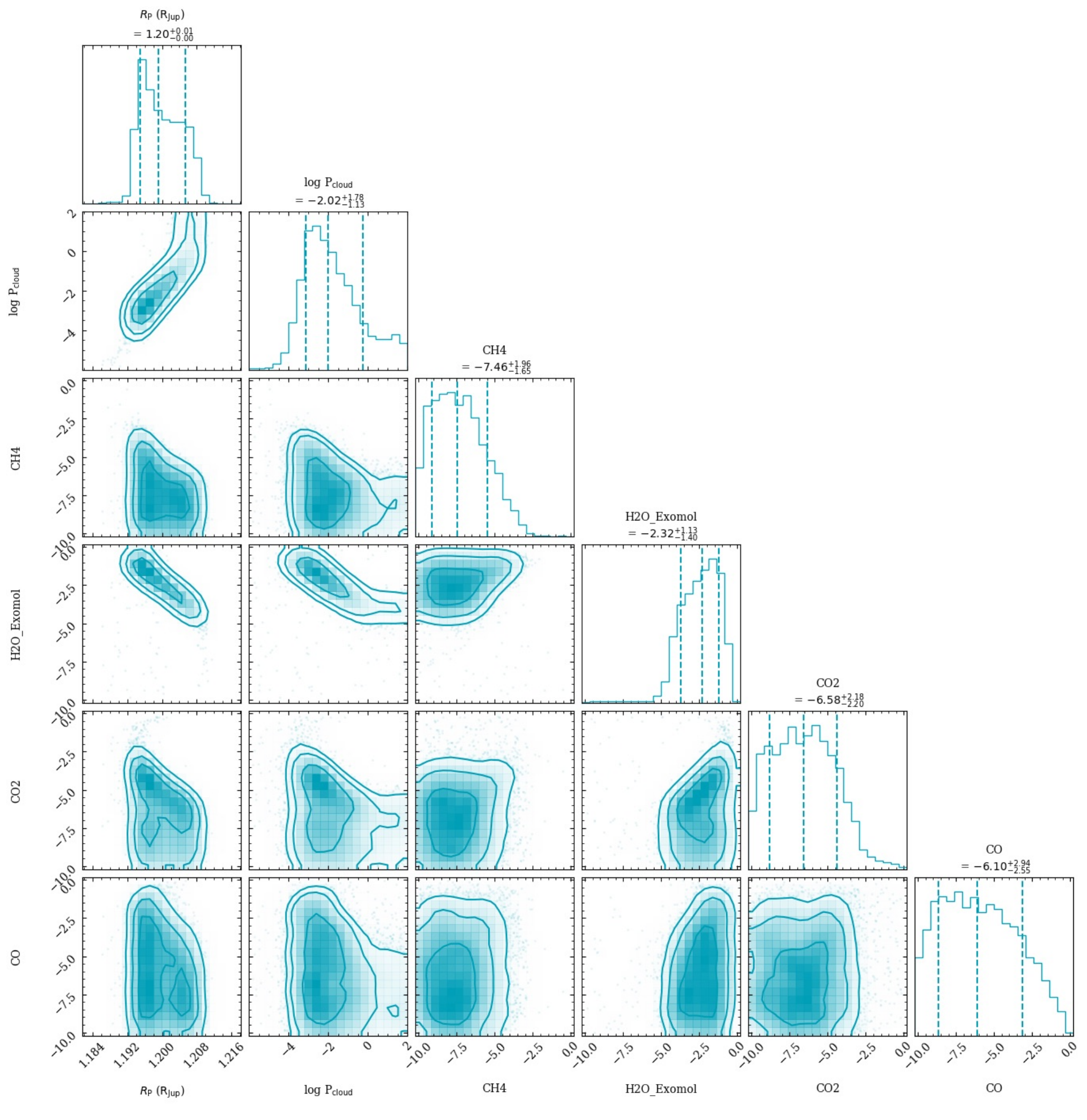}
    \caption{The posterior distribution of model atmosphere parameters for HAT-P-14 b, assuming isothermal temperature and free chemistry. (both $P_{cloud}$ and the mass fractions of atmosphere components are in $\log_{10}$ scale)}
    \label{fig:posterior}
\end{figure}

\begin{figure}
    \centering
    \includegraphics[width=1\columnwidth]{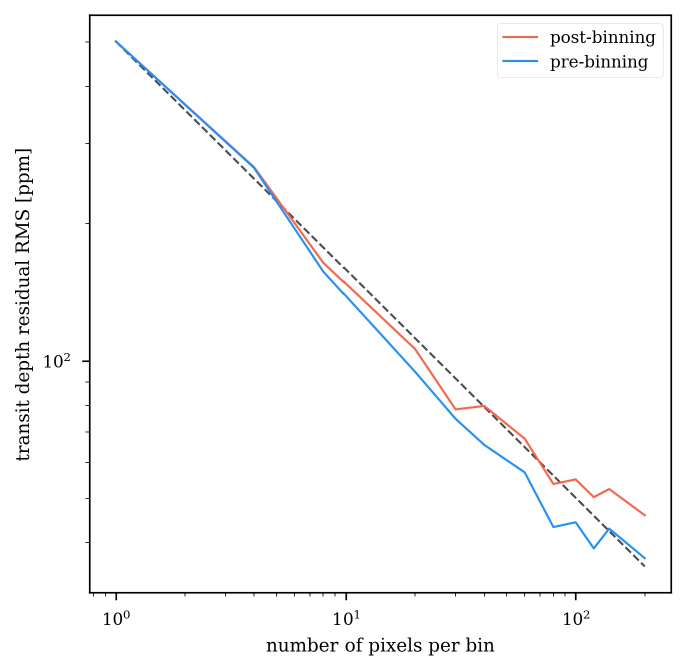}
    \caption{the residual Root Mean Square of transit depth calculated with different numbers of pixels per wavelength bin, from binning before (pre-binning) or after (post-binning) the light curve fit. The dashed black line is the theoretical decrease for RMS (proportional to $1/\sqrt{N}$)}
    \label{bin}
\end{figure}


We test and compare two different wavelength binning methods to calculate the transmission spectrum. The first way is to perform the spectral binning of the light curve before fitting the transmission spectrum. This ``pre-binning'' is done by setting a minimum number of pixels per bin and then dividing the wavelength bin accordingly. The second way is to first fitting the transmission spectrum at the native resolution of the instrument and then, at the end, taking the weighted average of the transit depth in each wavelength bin to obtain the binned transmission spectrum (``post-binning"). We compare the two methods by calculating the residual Root Mean Square (RMS) of the transit depth obtained by two binning methods, with different numbers of pixels per bin. If the transmission spectra is a featureless flat line, the transit depth RMS would be an empirical measurement of the total noise properties in TSO data when extracting a transmission spectra.  

From Figure \ref{bin}, we can see that the residual RMS of transit depth is lower if we do the binning before light curve fit and higher if we do the binning in post-processing. The results indicate that it is better to do ``pre-binning'', i.e. first divide the light curve to wavelength bins and then perform transmission spectrum extraction by fitting the transit depth together with other parameters. 

However, we should consider that the existence of an atmosphere could complicate the interpretation of our results. If the real transmission spectrum is not a a perfect flat line, the residual RMS would not as informative. We encourage this test to be repeated on a more optimal calibration target which has a higher mass than HAT-P-14 b and hence a more featureless spectrum and in a less grazing transit geometry. Repeating the test for a flat spectrum will give us a better comparison between different binning methods.


\subsection{Atmospheric Retrieval Setup}
\label{sec:setup}

We use the open-source Python package PetitRADTRANS (pRT, \cite{Molli_re_2019}) to model the atmosphere of planet HAT-P-14 b. pRT is a package designed to generate transmission and emission spectra of exoplanets that can work at both high ($\lambda/\Delta \lambda = 10^6$) and low resolution ($\lambda/\Delta \lambda = 1000$) for clear or cloudy atmospheres. In our study, we use it to generate the model transmission spectra of HAT-P-14 b. We assume a 1D atmosphere with an isothermal temperature profile ($T_{\text{iso}} = 1624K$, \citealt{2012MNRAS.426.1291S}). We use a fixed value of surface gravity $\log_{10}g =3.6262$ calculated from $M_p = 3.44 M_J$, $R_p = 1.42 R_J$ \citep{2017AJ....153..136S}. We adopt the free chemistry model by letting the chemical abundances of different gases vary freely within boundaries but assuming the abundances to be constant vertically in the atmosphere. The default opacity database included in pRT is used as input. When defining the model, We include the line opacities of $\mathrm{CO_2}$, $\mathrm{CO}$, $\mathrm{H_2O}$, and $\mathrm{CH_4}$. The rest of the atmosphere is filled by $\mathrm{H_2}$ and $\mathrm{He}$ \textbf{in a solar-abundance ratio of $0.766:0.234$}. The Rayleigh scattering of $\mathrm{H_2}$ and $\mathrm{He}$ and the $\mathrm{H_2}$-$\mathrm{H_2}$, $\mathrm{H_2}$-$\mathrm{He}$ collision-induced absorption (CIA) opacities are taken into account. For each line species, we use a log-uniform prior between $10^{-10}$ and $1$. We also include a grey cloud deck in the model, parameterized by its base pressure $P_{\text{Cloud}}$.

The pRT package uses the correlated-k method for computing opacities at model resolution $\lambda/\Delta \lambda = 200$, and the model spectrum is then binned down to the same resolution as our observed spectrum. We fit the model with the nested sampling algorithm (\cite{10.1063/1.1835238}) using the Python package PyMultiNest (\cite{2009MNRAS.398.1601F}, \cite{2014A&A...564A.125B}) to determine the posterior abundances of possible gas species in the atmosphere.  In our initial run used for determining the offset between SOSS and NIRSpec, there are 7 free parameters in the atmosphere retrieval: planet radius ($R_p$), gray cloud base pressure ($P_{cloud}$), the mass fractions of four gas species -- $\mathrm{CO_2}$, $\mathrm{CO}$, $\mathrm{H_2O}$, $\mathrm{CH_4}$, and the spectrum offset for SOSS. The initial run gives an offset of 237ppm between SOSS and NIRSpec. For the rest of the retrieval run, we first add the offset to SOSS part of the spectrum and then run retrieval on the full, pre-offsetted transmission spectrum with 6 free parameters. We use 2000 live points for each MultiNest sampling process. 


\subsection{Atmosphere of HAT-P-14 b}

 The retrieved mass fractions of gas species are shown in Table \ref{tab:posteriors}. The resulting best-fit model spectrum and the observed spectrum, the residual between the two, and the major contributing atmosphere components are shown in Figure \ref{fig: comb_spec}. The corner plot for the posterior distribution of the atmosphere parameters is shown in Figure \ref{fig:posterior}. From the posterior distribution, we can see that the retrieved mass fraction of $\mathrm{H_2O}$ ($\log X_{\mathrm{H_2O}}=-2.32^{+1.13}_{-1.40}$) are better constrained than the mass fractions of $\mathrm{CO_2}$($\log X_{\text{CO$_2$}}=-6.58^{+2.18}_{-2.20}$), $\mathrm{CO}$($\log X_{\text{CO}}=-6.10^{+2.94}_{-2.55}$), and $\mathrm{CH_4}$($\log X_{\text{CH$_4$}} = -7.46^{+1.96}_{-1.65}$). It can be seen in Figure \ref{fig: comb_spec} that the small atmospheric bumps  ($\sim 100$ppm, occurring near $1.4\mu m$, $1.9\mu m$, $2.6\mu m$ in observed transit spectrum aligns with the features of water vapor well, suggesting the existence of water vapor is likely to be real. Also, the best-fit model gives a reduced chi-square value of $\chi_\nu = 0.98$ (DOF= 194) as compared to a flat line $\chi_\nu = 1.10$ (DOF =199), suggesting that the model with atmosphere provides better fit to the observed data. 

We determine the significance of gas species contribution by Bayesian model comparison. The comparison is done by repeating the previously mentioned free chemistry retrieval after removing each gas species or cloud deck. We compare the marginal evidence of the model with one parameter removed ($\ln Z_s$) to the evidence of the full model ($\ln Z_{\text{full}}$ to determine the significance of each parameter. The results of this model comparison are presented in Table \ref{tab:model_comparison}. We use the $\ln B_{10} = |\Delta \ln Z |$ values of $1.0, 2.5, 5.0$ as the thresholds of evidence to be ``weak", ``moderate" and ``strong" for comparing two models $\mathcal{M}_0$ versus $\mathcal{M}_1$ \citep{Trotta_2008}. Among the 5 parameters, $\mathrm{H_2O}$ is moderately favored ($+3.09\sigma$) and is the only favored gas species; the cloud deck is weakly favored ($+1.90\sigma$); while $\mathrm{CO_2}(-1.83\sigma)$, $\mathrm{CO}(-1.75\sigma)$ and $\mathrm{CH_4}(-1.96\sigma)$ are all weakly disfavored. Therefore, we claim possible detection of $\mathrm{H_2O}$ and non-detection of other gas species. For the non-detected species, we constrain its upper bound on the mass fraction by taking the 95th percentile of its distribution. The obtained upper bound are respectively $\log X_{\mathrm{CO_2}}^{\text{upper}} = -3.23$, $\log X_{\mathrm{CO}}^{\text{upper}} = -1.69$, and $\log X_{\mathrm{CH_4}}^{\text{upper}} = -4.36$. 

We also run a free-temperature retrieval to test the accuracy of the atmospheric abundances. The retrieved temperature value, $T_\text{iso} = 2324 $K, is higher than the fixed temperature ($T_\text{iso} = $1624K) used in the retrieval. The free-temperature model is weakly favored ($+1.49\sigma$) compared to the fixed-temperature model, but they give consistent estimates for the atmospheric parameters: the mass fractions given by the free-temperature retrieval are $\log X_{\mathrm{H_2O}}=-2.32^{+1.06}_{-1.41}$, $\log X_{\text{CO$_2$}}=-7.00^{+2.27}_{-1.95}$, $\log X_{\text{CO}}=-6.21^{+2.83}_{-2.54}$, $\log X_{\text{CH$_4$}} = -7.14^{+2.20}_{-1.87}$.

\subsection{Equilibrium Chemistry Retrieval}
We also run a retrieval with equilibrium chemistry to better constrain the metallicity of the atmosphere of HAT-P-14 b. For this retrieval, we follow the same model setup with same line species and filler species as described in Section \ref{sec:setup}. Instead of letting the mass fractions of line species vary freely, we assume the species are in chemical equilibrium. There are 4 free parameters in the retrieval: planet radius ($R_p$), gray cloud base pressure  ($P_\text{{cloud}}$), metallicity ($\mathrm{[Fe/H]}$), C to O ratio ($\mathrm{C/O}$) and $\mathrm{P_{quench}}$,  the pressure below which the abundances of $\mathrm{H_2O}$, $\mathrm{CH_4}$ and CO are taken to be constant. The retrieved parameters with equilibrium chemistry is shown in Table \ref{tab:equil}. The retrieved metallicity is $\mathrm{[Fe/H]}=-0.08^{+0.89}_{-0.98}$, and C to O ratio is $\mathrm{C/O}=0.41^{+0.24}_{-0.20}$. The equilibrium chemistry model gives a tighter constraint on metallicity and we found the metallicity to be close to solar value.  
The measured C/O ratio indicates an oxygen-rich atmosphere, which is consistent with our detection of $\mathrm{H_2O}$. 
\subsection{M-dwarf Companion}

\begin{figure*}[t]
   \plottwo{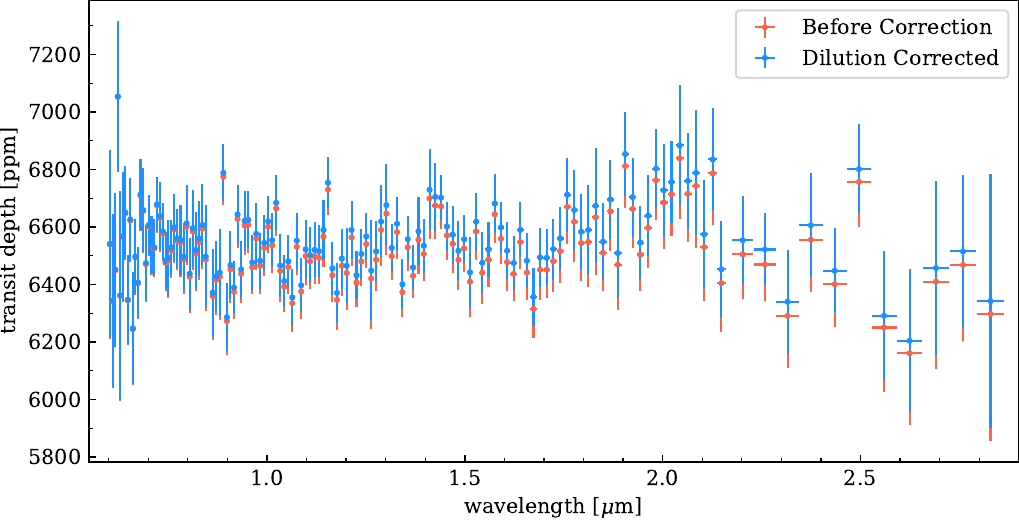}{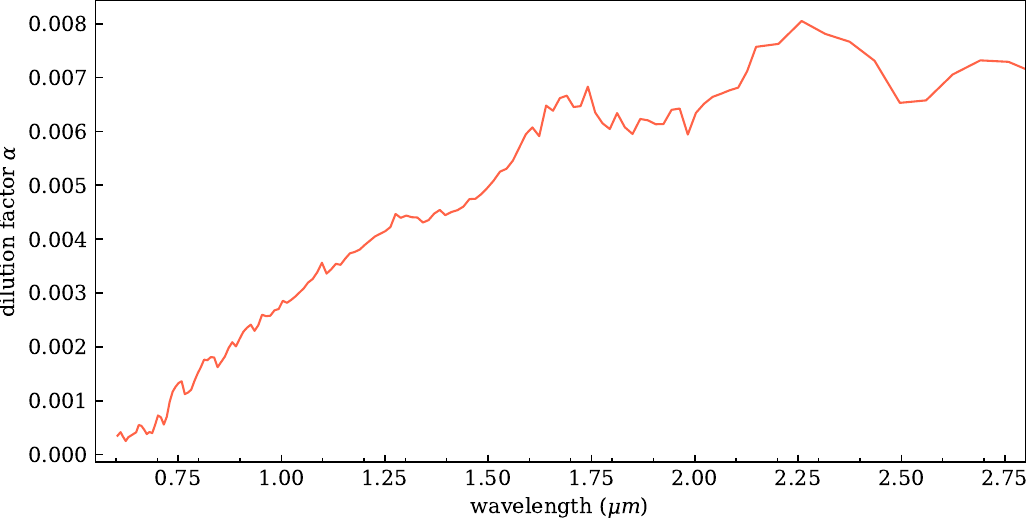}
   \caption{\textit{Left}: Dilution corrected SOSS transmission spectrum overplotted on the original transmission spectrum. \textit{Right}: The maximum possible dilution factor introduced to the transmission spectrum from the nearby M-dwarf companion from 0.6 $\mu m$ to 2.8 $\mu m$. The spectrum is generally unaffected by the possible dilution; in particular, the water bands are still present in the spectrum after the correction.}
   \label{fig:dilution}
\end{figure*}

As shown by A23, a companion star bound to HAT-P-14 b is seen in the 64 $\times$ 64 SOSS acquisition subarray and has the color of an early M-type star. This could cause dilution in the region where spectral dispersions are overlapped. Moreover, since M-dwarfs harbor water bands, the contamination could mimic the water signal in the transmission spectra. To investigate whether the water feature is from the companion star, we estimate the maximum degree of dilution that can be caused by the companion star by considering the worst-case scenario in which the spectral traces from the target star and companion star overlap completely. We normalize the throughput-weighted 6600 K (corresponding to the target star) and 3300 K (corresponding to the companion star) PHOENIX models to the $\Delta$F480M magnitude ($\Delta$F480M $\approx ~5$ mag) seen at acquisition. The dilution-corrected transit depth $\delta'$ is then estimated by
\begin{align}
    \delta' &= \delta \cdot (1+\alpha)
\end{align}
where $\delta$ is the measured transit depth and $\alpha$ is the dilution factor from the companion star, which is simply the flux ratio between the companion star and the target star returned by the normalized PHOENIX models. We find in the worst case, the contamination from the companion star does not produce a signal strong enough to mimic the water feature in our observed transmission spectrum. After correcting the dilution, the water features at 1.4 $\mu m$ and 1.9 $\mu m$ are still robustly present. Therefore it is safe for us to omit the possible dilution from the companion star. We show the dilution-corrected transmission spectrum and the maximum dilution factor in Figure \ref{fig:dilution}.

\begin{deluxetable}{lCC}
\label{tab:posteriors}
\tablecaption{Parameter Estimation for Free Chemistry Retrievals}
\tablewidth{0pt}
\tablehead{
\colhead{Parameters}& \colhead{Prior}  &  \colhead{\begin{tabular}{c} Posterior \\ ($1\sigma$ uncertainty) \end{tabular}}
}
\startdata 
$\log X_{\text{H$_2$O}}$ & $\mathcal{U}(-10,0)$& $-2.32^{+1.13}_{-1.40}$ \\
$\log X_{\text{CO$_2$}}$ &$\mathcal{U}(-10,0)$ &$-6.58^{+2.18}_{-2.20}$ \\
$\log X_{\text{CO}}$ &$\mathcal{U}(-10,0)$ &$-6.10^{+2.94}_{-2.55}$\\
$\log X_{\text{CH$_4$}}$ & $\mathcal{U}(-10,0)$&$-7.46^{+1.96}_{-1.65}$ \\
$\log (P_{\text{cloud}})$ (bars)& $\mathcal{U}(-6,2)$ &$-2.02^{+1.78}_{-1.13}$ \\
\enddata
\end{deluxetable}

\begin{deluxetable}{lCCl}
\label{species}
\tablecaption{Model comparison for atmospheric components. 
 $\Delta \ln Z = \ln Z_{\text{full}} - \ln Z_{s}$, where $\ln Z_{\text{full}}$ is the marginal evidence for the full free chemistry model and $\ln Z_{s}$ is the marginal evidence for the model without gas species $s$}

\label{tab:model_comparison}
\tablewidth{0pt}
\tablehead{
\colhead{Species} & \colhead{$\Delta \ln Z$} & \colhead{Significance} & \colhead{Inferences}
}
\startdata 
H$_2$O  &$+3.4$ & $+3.09~$\sigma & Moderately favored \\
CO$_2$  &$-0.7$ &$ -1.83~$\sigma & Weakly disfavored \\
CO  &$-0.6$ & $-1.75~$\sigma & Weakly disfavored \\
CH$_4$  & $-0.9$ & $-1.96~$\sigma & Weakly disfavored \\
Cloud & $+0.8$ & $+1.90~$\sigma & Weakly favored \\
\enddata
\end{deluxetable}

\begin{deluxetable}{lCC}
\label{tab:equil}
\tablecaption{Parameter Estimation for Equilibrium Chemistry Retrievals}
\tablewidth{0pt}
\tablehead{
\colhead{Parameters}& \colhead{Prior}  &  \colhead{\begin{tabular}{c} Posterior \\ ($1\sigma$ uncertainty) \end{tabular}}
}
\startdata 
$\log (P_{\text{cloud}})$ (bars)& $\mathcal{U}(-6.0,2.0)$ &$-2.02^{+1.78}_{-1.13}$ \\
$\mathrm{[Fe/H]}$ & $\mathcal{U}(-1.5,1.5)$ &$-0.08^{+0.89}_{-0.98}$ \\
$\mathrm{[C/O]}$ & $\mathcal{U}(0.1,1.6)$ & $-0.41^{+0.24}_{-0.20}$  \\
$\log (P_{\text{quench}})$ & $\mathcal{U}(-6.0,3.0)$ & $-2.84^{+1.98}_{-2.07}$  
\enddata
\end{deluxetable}

\section{Discussion}\label{sect:discussion}

The transmission spectrum presented in this work has different implications from the previously published work (A23, E23). In particular, the NIRISS/SOSS spectrum from 0.6 $\mu m$ to 2.8 $\mu m$ presented in this work does not exhibit a slope at the blue end reported by A23. The sudden increase of measured opacity at 1.4 $\mu m$ is also not seen. We speculate that two factors could be causing these differences: 
\begin{enumerate}
    \item The spectrum presented by A23 is processed from fitting spectroscopic light curves at native levels first and then binned afterward. This could cause the systematics underneath to be poorly constrained due to lower signals at native-resolution light curves, and thus lead to differences in measured transmission spectra (See Schmidt $\&$ Tsai et al in prep). However, we also do not see the same features in the native-resolution transmission spectra, so this factor could be reasonably ruled out. 
    \item The version of the STScI \verb|jwst| pipeline used in this work is different from A23. Many improvements for the \verb|jwst| pipeline, in particular the superbias subtraction step for the NIRISS/SOSS, have been made since A23. These improvements can lead to different transmission spectra compared to those processed from older versions of \verb|jwst| pipeline. For example, the first transmission spectrum of WASP-96b with NIRISS/SOSS also features a blueward slope \citep{2023MNRAS.524..835R} that was not seen by Hubble at the same wavelengths. However, later analysis of the same dataset with the latest \verb|jwst| pipeline reported non-detection of such slope (Wang et al in prep). This highlights the need to reanalyze the early NIRISS/SOSS transiting exoplanet targets with the latest methods.
\end{enumerate}


The presence of H$_2$O in the atmosphere of HAT-P-14 b is consistent with the equilibrium chemistry for this type of planet. Though a flat line can not be completely excluded as it fits the transmission spectrum well ($\chi^2_{\nu} = 1.10$, dof $ = 199$), the measured transmission spectrum does not appear to be completely featureless: that the presence of an atmosphere with small bumps that line up with H$_2$O bands improves the fit are indications that they are real atmospheric features rather than unconstrained systematics. In addition, these water bands are seen in data from two independent instruments (both across the SOSS wavelength range and in the bluest wavelengths of G395H), which adds to the robustness of the detection. Since HAT-P-14 b is massive (3.44 $M_J$), it would be expected to have low metallicity. The non-detection of CO$_2$ is consistent with the  mass-metallicity relationship \citep{2016ApJ...831...64T}.

\textbf{We consider the possibility that the detected H$_2$O spectral feature could be caused by stellar contamination through the transit light source effect (TLSE, \citealt{2018ApJ...853..122R}). This effect occurs when unocculted stellar spots produce spectral features similar to planetary atmospheric features. However, HAT-P-14 is known to be an inactive star, as indicated by no detectable emission in the Ca H and K lines \citep{2010ApJ...715..458T}.  Spectral modeling has shown that F-type stars typically has minimal rotational variability due to stellar spots \citep{2019AJ....157...96R}. In addition, stellar spots on these F-stars do not reach the low temperatures required for showing significant H$_2$O absorption features. Also, the photospheric light curve measured by TESS does not exhibit any obvious sinusoidal periodicity indicative of rotating stellar spots on HAT-P-14. We therefore conclude that the detected H$_2$O absorption features are planetary in nature. }

HAT-P-14 b has a Transmission Spectroscopy Metric (TSM, \citealt{2018PASP..130k4401K}) of 23.64, placing it the 805th best target for atmospheric characterization with transmission spectroscopy. The constrained atmosphere of HAT-P-14 b with combined NIRISS and NIRSpec analysis shown in this work indicates JWST's capability to characterize atmospheres of nearly a thousand of exoplanets. It also shows the promise of detecting the secondary atmospheres of terrestrial planets with JWST, signals with amplitudes of tens of ppm like the case of HAT-P-14 b.
    
\section{Conclusion}\label{sec:conclusion}
In this work, we present the combined JWST NIRISS SOSS and NIRSpec G395H transmission spectroscopy of the super-Jupiter HAT-P-14 b, one of the JWST commissioning targets. While HAT-P-14 b is placed as the 805th best planet for transmission spectroscopy because of its small atmospheric scale height and hence featureless expected spectrum, we detect small water features and gray cloud deck in the atmosphere of HAT-P-14 b consistently across the two independent JWST instruments.  This is a completely different interpretation from the initial results reported by A23 and E23, and we suspect it is primarily caused by the improved STScI \verb|jwst| pipeline, as also shown by Wang et al. in prep. Therefore, this work highlights the need to reanalyze the early JWST NIRISS SOSS transiting exoplanet targets. The detection of water is consistent with equilibrium chemistry, which predicts water features of tens of ppm. That we are able to detect water and constrain multiple other species (CO$_2$, CO, and CH$_4$) for HAT-P-14 b showcases the unique and unparalleled capability of JWST to characterize nearly a thousand exoplanets through transmission spectroscopy.

\bibliography{main}{}
\bibliographystyle{aasjournal}



\end{document}